# Scalable cyclic transformation of orbital angular momentum modes based on a nonreciprocal Mach-Zehnder interferometer

Yu-Fang Yang[1], Ming-Yuan Chen[1], Feng-Pei Li[1], Ya-Ping Ruan[1], Zhi-Xiang Li[1], Min Xiao[1, 2], Han Zhang[1, 3, 4], and Ke-Yu Xia[1, 3, 5]

[1]*National Laboratory of Solid State Microstructures, School of Physics, College of Engineering and Applied Sciences, and Collaborative Innovation Center of Advanced Microstructures, Nanjing University, Nanjing 210093, China*
[2]*Department of Physics, University of Arkansas, Fayetteville, Arkansas 72701, USA*
[3]*Hefei National Laboratory, Hefei 230088, China*
[4]*email: zhanghan@nju.edu.cn*
[5]*email: keyu.xia@nju.edu.cn*

**Abstract:** The orbital angular momentum (OAM) of photons provides a pivotal resource for carrying out high-dimensional classical and quantum information processing due to its unique discrete high-dimensional nature. The cyclic transformation of a set of orthogonal OAM modes is an essential building block for universal high-dimensional information processing. Its realization in the quantum domain is the universal quantum Pauli-X gate. In this work, we experimentally demonstrate a cyclic transformation of six OAM modes with an averaged efficiency higher than 96% by exploiting a nonreciprocal Mach-Zehnder interferometer. Our system is simple and can, in principle, be scaled to more modes. By improving phase stabilization and inputting quantum photonic states, this method can perform universal single-photon quantum Pauli-X gate, thus paving the way for scalable high-dimensional quantum computation.

## 1. Introduction

It is well known that the Laguerre-Gaussian (LG) mode of a light beam carries a helical phase term of $\exp(il\theta)$, where $\theta$ is the azimuthal angle, and $l$ the topological charge, indicating the orbital angular momentum (OAM) of photons. Each photon has a well-defined OAM of $l\hbar$ [1]. The OAM of photons provides high discrete dimensions for encoding information. Thus, it becomes a promising candidate for high-dimensional information processing. For this reason, it has been widely used in numerous classical [2-4] and quantum communication protocols [5-8], two-photon quantum entanglement in a high-dimensional Hilbert space [9-13], and multi-photon entanglement [14-16].

Unitary transformations leveraging the OAM of photons play the central roles in high-dimensional information technologies. Classical transformation between high-order LG and HG modes has been performed by correctly arranging the cylindrical lenses [17]. Recently, a classical four-mode cyclic transformation has been first realized by using OAM beam splitter (OAM-BS) modules [18], which are the Mach-Zehnder interferometers (MZIs) built from beam splitters (BSs) and Dove prisms. The principle of classical cyclic transformation has been extended to accomplish a four-dimensional Pauli-X gate by adding a phase-locking system and utilizing quantum tomography technique [19]. Subsequently, a five-dimensional Pauli-X gate is implemented by modulating the wavefront of OAM beam through the wavefront matching technique [20]. Multi-mode cyclic transformation based on the OAM beams is the basis for the high-dimensional Pauli-X gates. Arbitrary unitary transformation for single-photons can be achieved by the combination of high-dimensional Pauli-X and Pauli-Z gates [21]. The high-dimensional Pauli-Z gate is simply realized by using a Dove prism [22]. However, both classical cyclic transformation and quantum Pauli-X gates are limited up to five modes thus far. It is highly desirable to extend unitary transformations to more modes in the OAM-mode basis.

To leverage the OAM of photons for high-dimensional information processing, it is crucial to spatially separate the OAM modes. A variety of schemes are proposed to accomplish this task [23-30]. Among these methods, the OAM-BS module achieves a great success. It can separate the odd and even OAM modes [23, 24]. Zhang et al. improve this OAM-BS module by replacing the BSs with polarization beam splitters (PBSs) and separate the OAM modes [25]. To apply this module to a large number of modes, multiple MZIs is needed to be cascaded [26]. This greatly increases the experimental complexity and reduces the reliability and the fidelity. It remains a challenge to conduct many-mode unitary transformations without considerably adding the system complexity. The FP cavity have been used to sorting the OAM modes [30]. It is yet to be exploited for high-dimensional unitary transformations.

In this work, we experimentally realize a six-mode cyclic transformation with an efficiency larger than 96% by building a nonreciprocal MZI. This MZI leverages the nonreciprocal control of the optical circulator and as an OAM filter.

The paper is organized as follows: In Section 2, we explain the concept and the system for the high-dimensional cyclic transformation based on photonic OAMs. Then, the experimental setup is shown in Section 3. In Section 4, we demonstrate the performance of the proposed cyclic transformation by using six OAM modes as an example. In the end, we summarize the results and discuss the potential applications of our scheme.

## 2. System and Concept

A classical cyclic transformation and the quantum Pauli-X gate encoded in a high-dimensional OAM basis perform the conversion of $(|-(l+1)\rangle, |-l\rangle, ..., |l\rangle) \rightarrow (|-l\rangle, |-l+1\rangle, ..., |l+1\rangle)$. Fig. 1(a) conceptually depicted our method using a nonreciprocal MZI for implementing a high-dimensional cyclic transformation and Pauli-X gate. The nonreciprocal MZI consists of two optical circulators, two identical FP cavities and an OAM flipper. To conduct cyclic transformation, we input a vortex OAM mode of light beam for the cyclic transformation or the single-photon superposition state of OAM modes for the Pauli-X gate. A Spiral phase plate (SPP) is used to convert the $l$th OAM mode of the input field to the $(l+1)$th mode. As demonstrated in [30], the resonance frequency of a FP cavity is crucially dependent on the absolute of the topological charge $|l|$ of the OAM mode. The key idea of our method leverages this $l$-dependent nature of a FP cavity and the nonreciprocal light modulation of an optical circulator. In our arrangement, the FP cavity is resonant with the highest-order $(l+1)$th mode, but largely detuned with other modes. It transmits the on-resonance $(l+1)$th mode but reflects other off-resonance modes. In ideal case, the spectrum of the input field is well within the linewidth of the FP cavity. The transmission and reflection are nearly unity, see Fig. 1(b). however, the transmitted and reflected fields keeps their phases unchanged, see Fig. 1(c). With the first pair of the circulator (C1) and the FP cavity (FP1), the reflected modes are led to the second circulator (C2). The OAM of the transmitted field of the FP1 is first flipped and then incidents to the second FP cavity (FP2), which is assumed identical with the FP1. This flipper is used for converting the transmitted $(l+1)$th mode to the $-(l+1)$th input mode. This flipper can also be inserted into another arm of the MZI. Then, the field transmits the FP2. The modes reflected off the FP1 are input to the FP2 via the C2 and then reflected again by the FP2. The overall reflected fields and the transmitted field is combined with the C2 as the final output. Thus, we can perform the high-dimensional classical cyclic transformation and the quantum Pauli-X gate with this nonreciprocal MZI forming with these two pairs of the circulators and the FP cavities.

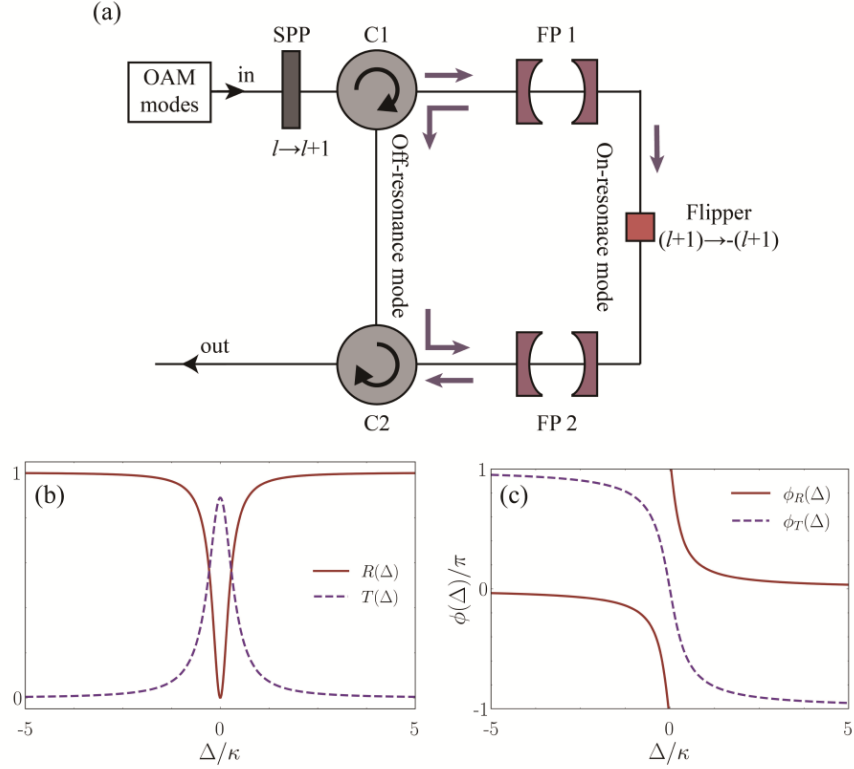

Fig. 1 (a) The conceptual diagram for cyclic transformation of OAM modes and the high-dimensional Pauli-X gate based on a nonreciprocal MZI. A SPP increases the topological charge of an OAM mode by one to the input $l$th mode. Optical circulators are used to redirect light reflected off the FP cavity to the third path. The FP1 and FP2 cavities are identical. They transmit the selected OAM mode but reflect other modes. A flipper is used to convert the transmitted $(l+1)$th mode to the $-(l+1)$th one. (b) The reflection and transmission spectral of the FP cavity. (c) The phases of the reflected and transmitted fields.

Below we provide an explanation for the $l$-dependent transmission and reflection of the FP cavity. The transverse distribution of the LG mode in the FP cavity is given in cylindrical coordinates under the paraxial approximation by [1],

$$u_{pl}(r,\theta,z) = \frac{C}{\sqrt{1+z^2/z_R^2}}\left[\frac{\sqrt{2}r}{w(z)}\right]^l L_p^l\left[\frac{2r^2}{w(z)^2}\right]\exp\left[\frac{-r^2}{w(z)^2}\right] \\ \times \exp\left[\frac{-ikr^2 z}{2(z^2+z_R^2)}\right]\exp(-il\theta)\exp[i(2p+|l|+1)\psi(z)]. \quad (1)$$

In the above equation, $C$ is a normalization constant; $z_R$ is the Rayleigh range of the beam; $w(z)$ is the radius of the beam; $L_p^l$ is the generalized Laguerre-Gauss polynomial with azimuthal and radial mode indices $l$ and $p$, respectively. The term $\psi(z) = \tan^{-1}(z/z_R)$ is the Gouy phase. The above equation implies that higher-order LG modes have an additional Gouy phase shift with respect to Gaussian beam. This Gouy phase is attributed to the increased transverse structure of the beam. It enables us to separate out the different OAM modes by using the FP cavities. Based on this phase shift and the standing wave condition of the cavity, the resonance condition of the FP cavity can be written as [30],

$$\omega = \frac{\pi c}{D}\left[q + (2p + |l| + 1)\frac{\phi}{\pi}\right], \qquad (2)$$

where $D$ is the distance between the front and rear mirrors of the FP cavity, $q$ is an integer, and the accumulated Gouy phase shift of the LG modes is $\phi = \arccos\left(\pm\sqrt{(1 - D/R_1)(1 - D/R_2)}\right)$ when the light beam travels from one end of the cavity to the other. It is determined by the mirror curvatures ($R_1$ and $R_2$). For a fixed cavity length $D$, the LG modes clearly exhibit distinct resonant frequencies dependent on $|l|$.

Exploiting the Gouy-phase-induced frequency splitting, we can filter a selective OAM mode with a FP cavity. For a two-sided FP cavity, the cavity reflection and transmission coefficients can be derived from the input-output relation

$$r(\Delta) = 1 - \frac{2\kappa_l}{i\Delta + \kappa}, \qquad (3a)$$

$$t(\Delta) = \frac{2\sqrt{\kappa_l \kappa_r}}{i\Delta + \kappa}. \qquad (3b)$$

Here, $\Delta = \omega - \omega_c$ is the detuning between the input LG modes and the cavity resonance frequency. The decays caused by the left and right mirror have the rates $\kappa_l$ and $\kappa_r$, respectively. According to our experiment, we can assume $\kappa_l = \kappa_r$. The total decay rate of the cavity is denoted as $\kappa \equiv \Delta\omega$. The reflection and transmission spectrum of the FP cavity can then be expressed as $T(\Delta) = |t(\Delta)|^2$ and $R(\Delta) = |r(\Delta)|^2$. They are shown in Fig. 1(b). The reflected and transmitted photons also surfer to an additional scattering phase shift, which are $\phi_R(\Delta) = \text{Arg}[r(\Delta)]$ and $\phi_T(\Delta) = \text{Arg}[t(\Delta)]$, respectively, where the function $\text{Arg}[x]$ means the phase of a complex number $x$. The phase shifts versus detuning are plotted in Fig. 1(c). It can be found from Figs. 1(b) and 1(c) that the selected resonant LG mode directly passes through the cavity with zero scattering phase, while other off-resonance OAM modes are completely reflected off the cavity without phase change. Thus, the resonant OAM mode is separated/filtered by the FP cavity. There is no additional phase introduced among the OAM modes. This nature of the transmission and reflection of the FP cavity meets the key requirement of a high-dimensional X-gate: The relative phase of an input state in superposition needs to remain unchanged in the ideal case.

Experimentally, we can choose the $(l + 1)$th OAM mode resonant with the FP cavity by adjusting the cavity length $D$. Other involved OAM modes reflect off the FP cavity and then are directed to another path.

## 3. Experimental Setup

Here, we give an experimental demonstration of six-mode cyclic transformation using the above system and concept. The experimental setup is depicted in Fig. 2. It is composed of three parts: OAM modes preparation, cyclic transformation and OAM modes measurement.

In the OAM modes preparation section, the incident laser beam is a horizontally polarized zero-order Gaussian beam with the wavelength locked at 794.9693 nm. The Q-plate (QP) is a birefringent waveplate with spatially varying optical axes orientation that can only modulate the phase of the beam [31]. A sandwich combination composed of two quarter-wave plates (QWP) and one QP is equivalent in principle to one SPP. It can transform zero-order Gaussian mode into arbitrary higher-order LG mode without changing the polarization state. After passing through the first QWP-QP-QWP combination, the Gaussian beam is transformed into OAM beam with an azimuthal mode index of $\{l = -3, -2, -1, 0, 1, 2\}$.

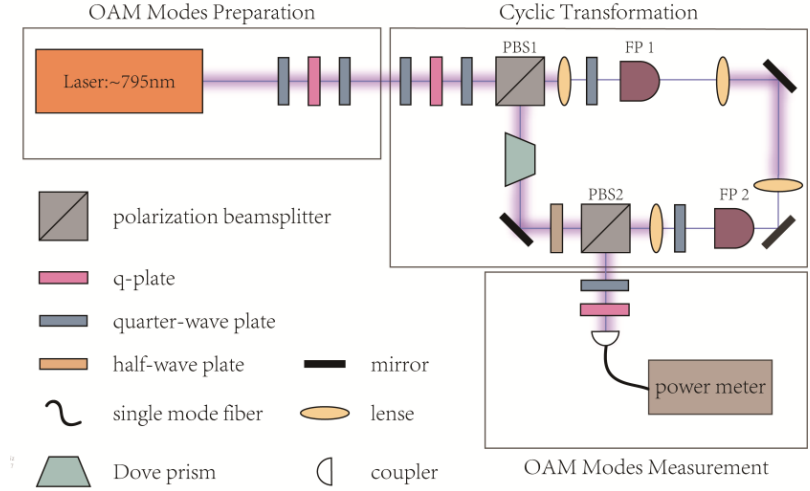

Fig. 2 Schematic of the experimental setup. The light source is a DL pro single-frequency continuous laser. A polarization beam-splitter (PBS) and a QWP form an optical circulator, which is used to redirect light reflected off the FP cavity to the third path [30]. The incident light is horizontally polarized. Light reflected by the FP1 (FP2) cavity passes the QWP twice, transforming into the vertically polarized light, and therefore exits at the reflective end of the PBS1 (PBS2). The thicker purple lines indicate the collimated laser beams.

After the OAM mode enters the cyclic transformation section, it goes through the second QWP-QP-QWP combination to shift to $(l + 1)$ mode, where $q = 1/2$ for QP. The OAM modes become the modes $\{l = -2, -1, 0, 1, 2, 3\}$. The FP1 and FP2 cavities are used as OAM modes separator and combiner, respectively. We experimentally adjust the length of the FP1 and FP2 cavities through temperature controllers so that the selected $l = 3$ mode is resonantly transmitted. Other modes with $l \neq 3$ are off-resonance with the cavities and thus reflected. The mode of $l = 3$ propagates in the right arm embedded with the FP cavities. It is reflected by two mirrors and then transmits the FP2 cavity. Finally, it exits the right arm of the MZI after reflected by the PBS2. Note that each reflection flips the OAM quantum number $l$. The 3th OAM mode is reflected odd (three) times. As a result, it changes to the mode with $l = -3$. Other OAM modes with $l \neq 3$ are reflected off the FP1 and FP2 cavities. The reflection of the cavity doesn't accumulate a phase shift to the OAM modes. We use the PBS and a quarter-wave plate to form an optical circulator. The lens right to the PBS is use for adjust the waist of the laser beams. The fields reflected off the FP1 is first directed by the first circulator into the left arm. A Dove prism is embedded into this arm and works as an OAM flipper. It adds one more OAM flipping in comparison to the right arm. The fields transmit the second circulator, are reflected off the FP2 and then redirected to the photodetector with the selected OAM mode from the right arm. After these processes, the input OAM modes $\{-3, -2, -1, 0, 1, 2\}$ are finally transformed to $\{-2, -1, 0, 1, 2, -3\}$ after passing through the cyclic transformation module. A cyclic transformation based on the OAM modes is realized.

In the OAM modes measurement section, a set of QWP and QP is used to convert the higher-order LG modes into the zero-order Gaussian mode, which are then collected through single-mode fiber and detected by power meter.

In general, for a vortex beam with $2l$ modes $\{-(l+1), -l, \ldots, l\}$, the cyclic transformation can be achieved by adjusting the lengths of FP1 and FP2 cavities so that they satisfy the resonance condition of the $(l + 1)$ mode. The maximum number of supported modes depends solely on the number of modes $l$ that the FP cavities can distinguish. This approach has a significant potential to be extended to higher dimensions.

## 4. Experimental Results and Discussion

In Equation (1), the electric field of the LG mode at its waist plane ($z = 0$) can be described as follows [32],

$$u_{p,l}(r,\theta) = \sqrt{\frac{2p!}{\pi w_0^2 (p+|l|)!}} \left(\frac{\sqrt{2}r}{w_0}\right)^{|l|} L_p^{|l|}\left(\frac{2r^2}{w_0^2}\right) \times exp\left(\frac{-r^2}{w_0^2}\right) exp(-il\theta). \qquad (4)$$

Where $w_0$ is the radius of the zero-order Gaussian beam at the waist. When the incident light beam is in the zero-order Gaussian mode, the QP can only modulate its phase to generate a vortex beam which is a special case of hypergeometric Gaussian beam [33]. Thus, the created beam is expressed as,

$$U_l(r,\theta) = exp\left(\frac{-r^2}{w_0^2}\right) exp(-il\theta). \qquad (5)$$

Hypergeometric Gaussian beam can be expanded as a superposition of LG modes with same topological charge $l$, and they can be written as,

$$U_l(r,\theta) = \sum_p C_p u_{p,l}(r,\theta), \qquad (6)$$

Here the probability amplitude $C_p$ is given by [32],

$$C_p = \sqrt{\frac{(p+|l|!)}{p!}} \frac{\Gamma\left(p+\frac{|l|}{2}\right)\Gamma\left(\frac{|l|}{2}+1\right)}{\Gamma\left(\frac{|l|}{2}\right)\Gamma(p+|l|+1)}, \qquad (7)$$

where $\Gamma(x)$ is the gamma function. The weightings of the fundamental radial mode ($p = 0$) decreases dramatically with increasing $l$ mode. The energy gradually shifts to higher-order radial modes ($p \neq 0$). The FP cavities have the identical resonance point for the same value of $(2p + |l|)$, but different values of $p, l$. This situation affects the accuracy of the cyclic transformation results. Thus, the weighting of the radial mode ($p = 0$) needs to be increased for a given higher-order $l$ mode. We use the lens to adjust the waist size of the incident light beam entering the FP1 cavity. When $w_0' = w_0/\sqrt{|l|+1}$, the energy content in the radial mode ($p = 0$) can be significantly improved, and in turn, the transmission of higher-order $l$ mode is enhanced [34]. For example, when the beam waist of the FP1 cavity is 50 $\mu m$, the beam waist of incident light beam with $l = \pm 3$ modes need to match the cavity beam waist, so that $w_0 = 50\ \mu m$. The lens is adjusted to reduce the beam waist of the incident vortex beam to $w_0' = 25\ \mu m$. As shown in Fig. 3(a), for two different beam waist sizes of incoming beam, the transmission of the FP1 cavity in different $l$ modes is measured. The measured data imply that decreasing the beam waist significantly increases the transmission of higher-order $l$ modes.

Experimentally, both FP1 and FP2 cavities are plano-convex lenses composed of monolithic fused silica crystal with 90% reflectance on both planar and convex surfaces. The convex surface of the lens has a curvature radius of 25mm. We employ temperature controllers to control the temperature of FP1 and FP2 cavities respectively. The temperature stability is less than 0.01 ℃ for 24 hours. Since the FP cavity is a single transparent crystal, it is not possible to scan the length of the FP cavity by attaching a piece of piezoelectric ceramic. Therefore, we place a photodetector behind the FP1 cavity and scan the wavelength of the laser to obtain a functional relationship between the transmission of the FP1 cavity and the wavelength of the laser, the experimental results are depicted in Fig. 3(b).

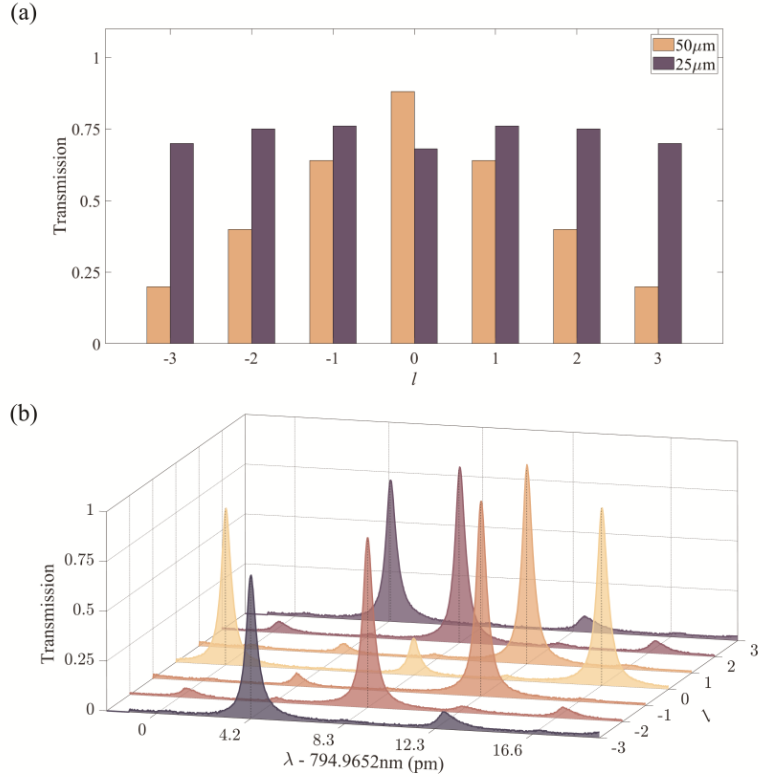

Fig. 3 Measured transmissions for the FP1. (a) Transmissions of the FP1 cavity for different $l$ modes for a 50 $\mu m$ and 25 $\mu m$ incident beam waist, respectively. (b) Transmission spectral of different $l$ modes through FP1. The wavelength interval between the two main peaks in the $l = 0$ mode means the free spectral range (FSR) of the cavity. The transmission spectral for paired modes $l = \pm 1$, $\pm 2$, and $\pm 3$ are almost identical because the FP1 cavity cannot distinguish the $+l$ and $-l$ modes.

Experimental data for FP1 cavity: free spectral range FSR = 7.90 GHz, bandwidth $\Delta\omega$ = 287 MHz, finesse $F = 27.4$, beam waist $w_0 = 50\ \mu m$. Theoretical data: FSR = 8.57 GHz, $\Delta\omega = 287$ MHz , $F = 29.9$, $w_0 = 55\ \mu m$ . The experimental and theoretical results are reasonably consistent. Experimentally, in order to increase transmission of FP1 cavity in the $l = \pm 3,\ p = 0$ modes, we reduced the beam waist of incident light beam to half of that of cavity mode ($w_0 = 50\ \mu m$) and observed a significant enhancement of transmission in the $l = \pm 3$ modes behind FP1 cavity. This also led to mode mismatch between the mode of incident beam and mode of FP1 cavity, for example, a small peak appeared in the middle of two large peaks when the incident beam is in the $l = 0$ mode. In addition, at the wavelength λ=794.9694 nm, the small transmission peaks appeared in the $l = \pm 1$ modes due to the presence of the $p = 1$ component in the vortex beams produced through the QP, which also satisfied the resonance condition of cavity in the $l = \pm 3, p = 0$ modes.

Due to experimental constraints, we are only able to modulate the phase of the incident light beam leading to the emergence of higher-order radial modes ($p \neq 0$) in the vortex beam. On the one hand, we narrow the incident beam waist to reduce the weighting of the higher-order radial modes. This method improves the transmission of the $p = 0$ mode but also generates mode mismatch between the input mode and the cavity mode; on the other hand, the generation of higher-order radial modes also lead to degeneracy of multiple OAM modes at the resonance point of FP cavity. If the phase and amplitude of the incoming light beam can be modulated simultaneously, for instance, by using SLM [35], it is anticipated to achieve a perfect OAM modes separation for FP cavities.

Our FP cavities can discriminate four cases for $\{|l| = 0,1,2,3\}$. Thus, the cyclic transformation device only works in six modes $\{l = -3, -2, -1, 0, 1, 2\}$. Experimentally we send each of the six OAM modes into the cyclic transformation device for testing one by one. For each input mode, each output mode after undergoing the cyclic transformation was performed projective measurement by a combination of QP and single-mode fiber, and a total of 36 measurements were completed. Fig. 4 (a) shows the experimental results, the distinct peaks imply that each input mode moved to the correct output mode after a cyclic transformation.

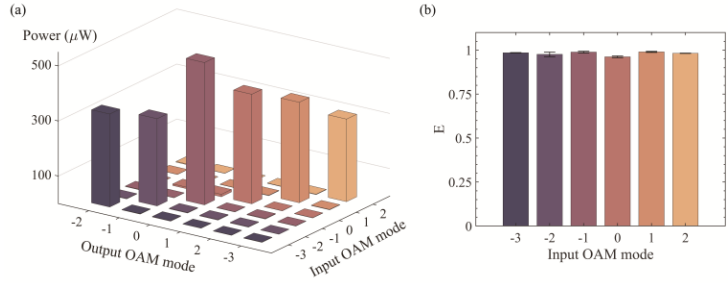

Fig. 4 Experimental results for six-mode cyclic transformation. (a) Measured power of six output OAM modes for six input OAM modes. Each color represents a specific input mode. (b) Measured efficiencies $E$ for a different input mode.

For a selected input mode, the efficiency $E$ of the correct output mode after the cyclic transformation is shown in Fig. 4 (b). The efficiency $E = I_c/I_{total}$ means the probability that the correct mode is created [18]. $I_c$ is the power of the correctly created mode and $I_{total} = \sum_{l=-3}^{+2} I_l$ is the sum of the intensity collected at all six collected output modes. The measured results clearly show that our six-mode cyclic transformation operation has a high efficiency, with an average efficiency greater than 96%.

The discrepancy of our measurement from the ideal cyclic transformation is mainly attributed to the fabrication imperfection of our FP cavities. The measured data implies that the FP cavity has a low transmission for the mode $l = 3$. The reason may be that FP cavities are flat-convex cavities. As a result, the transverse modes of the vortex beam from the convex side of the FP2 cavity are poor relative to that of planar side. This deformation of the transverse modes leads to a decrease in the collection efficiency. In addition, the coupling efficiency of the high-order modes is generally lower than that of the low-order modes in phase-flattening technology, which is also reflected in our experimental results [36].

## 5. Conclusions

By simply using the FP cavities, the optical circulators and a topological charge flipper, we have experimentally demonstrated the six-mode cyclic transformation with an average efficiency exceeding 96%. By further improving the fineness of the FP cavity and using an amplitude-phase modulator [35], it is feasible to achieve cyclic transformations for more OAM modes. Our method can also be straightforwardly used for conducting a high-dimensional quantum Pauli-X gate if the phase of the MZI can be stabilized at a level of about one of tens parts of the wavelength. Thus, this work provides a building block for high-dimensional quantum information processing.

**Funding.** This work was supported by National Key R&D Program of China (Grant Nos. 2019YFA0308700 and 2019YFA0308704), Innovation Program for Quantum Science and Technology (Grant No. 2021ZD0301400), the National Natural Science Foundation of China (Grant Nos. 11890704, 92365107, and 12305020), the Program for Innovative Talents and Teams in Jiangsu (Grant No. JSSCTD202138).

**Acknowledgments.** The authors thank Prof. Xi-Lin Wang, Pei Wan and Yi-Peng Zhang for their helpful discussions.

**Disclosures.** The authors declare that they have no conflict of interest.

**Data availability.** Data underlying the results presented in this paper are not publicly available at this time but may be obtained from the authors upon reasonable request.